\documentclass[11pt]{article}
\usepackage[usenames,dvipsnames,svgnames,table]{xcolor} % use color
\usepackage[obeyspaces,hyphens,spaces]{url}
\usepackage{jcapmod}
\usepackage{geometry}
\usepackage{changepage}

\usepackage{tabu}
\usepackage{booktabs}
\usepackage[english]{babel}
\usepackage{amsmath, amssymb, amsbsy, amstext, amsthm, simplewick}
\usepackage{hyperref}
\usepackage{graphicx}
\usepackage{amsfonts}
\usepackage{enumitem}
\usepackage{upgreek}
\usepackage{framed}
\usepackage{tensor}
\usepackage{pifont}
\usepackage{latexsym, mathrsfs}
\usepackage{array}
\usepackage{hyperref}
\usepackage{xspace}
\usepackage{longtable}
\usepackage{multirow}
\usepackage{cases}
\usepackage{empheq}
\usepackage{bm}
\usepackage{bbm}
\usepackage[table]{xcolor}
\usepackage[dvipsnames]{xcolor}
\usepackage{colortbl}
\usepackage{tablefootnote}
\usepackage{braket}
\usepackage[utf8]{inputenc}

\usepackage[load-configurations=astronomy, range-units=brackets, range-phrase=-, per-mode=reciprocal, mode=math]{siunitx}
\usepackage{threeparttable}
\usepackage{subfig}

\usepackage{ragged2e}

\usepackage[normalem]{ulem}

\usepackage{hhline}
\usepackage{array}
\newcolumntype{P}[1]{>{\centering\arraybackslash}p{#1}}
\usepackage{booktabs}
\usepackage{tikz}
\usetikzlibrary{calc,shadings,patterns,tikzmark,fadings}

\definecolor{Blue}{rgb}{0.25, 0.41, 0.88}
\definecolor{Red}{rgb}{0.92,0.,0.}
\definecolor{darkorange}{rgb}{1.0,0.549,0.}
\definecolor{cobalt}{RGB}{44, 98, 120}
\definecolor{Mathematica1}{rgb}{0.368417, 0.506779, 0.709798}
\definecolor{Mathematica2}{rgb}{0.880722, 0.611041, 0.142051}
\definecolor{Mathematica3}{rgb}{0.560181, 0.691569, 0.194885}
\definecolor{Mathematica4}{rgb}{0.922526, 0.385626, 0.209179}
\definecolor{Mathematica5}{rgb}{0.528488, 0.470624, 0.701351}
\definecolor{Mathematica6}{rgb}{0.772079, 0.431554, 0.102387}
\definecolor{Mathematica7}{rgb}{0.363898, 0.618501, 0.782349}
\definecolor{Mathematica8}{rgb}{1, 0.75, 0}
\definecolor{Mathematica9}{rgb}{0.647624, 0.37816, 0.614037}
\definecolor{plotBlue}{RGB}{94, 130, 181}
\definecolor{plotRed}{RGB}{233, 85, 54}
\definecolor{plotGreen}{RGB}{142, 176, 50}
\definecolor{plotPurple}{RGB}{135, 120, 178}

\renewcommand\_{\textunderscore\allowbreak}

\definecolor{cornellRed}{HTML}{B31B1B}
\definecolor{cornellBlue}{HTML}{0068AC}
\definecolor{cornellGreen}{HTML}{6EB43F}

\newcolumntype{C}[1]{>{\centering\let\newline\\\arraybackslash\hspace{0pt}}m{#1}}

\newcommand{\affsize}{10}

\makeatletter
\newlength{\apb@width}
\newcommand{\autoparbox}[2][c]{\settowidth{\apb@width}{#2}\parbox[#1]{\apb@width}{#2}}

\makeatother

\numberwithin{equation}{section}

\def\beq{\begin{equation}}
\def\eeq{\end{equation}}

\def\bea{\begin{eqnarray}}
\def\eea{\end{eqnarray}}

\def\beq{\begin{equation}}
\def\eeq{\end{equation}}
\def\bea{\begin{eqnarray}}
\def\eea{\end{eqnarray}}

\DeclareRobustCommand{\SkipTocEntry}[4]{}

\usepackage{colortbl}
\definecolor{blue2}{cmyk}{1, 0.1, 0.1, 0}

\definecolor{pyBlue}{RGB}{31, 119, 180}
\definecolor{pyRed}{RGB}{214, 39, 40}
\definecolor{pyGreen}{RGB}{44, 160, 44}
\definecolor{pyBlue2}{RGB}{0, 111, 237}
\definecolor{pyRed2}{RGB}{224, 52, 36}

\newcommand{\blue}[1]{\textcolor{pyBlue}{#1}}

\makeatletter
\def\Ddots{\mathinner{\mkern1mu\raise\p@
\vbox{\kern7\p@\hbox{.}}\mkern2mu
\raise4\p@\hbox{.}\mkern2mu\raise7\p@\hbox{.}\mkern1mu}}
\makeatother

\renewcommand{\frac}[2]{\dfrac{#1}{#2}}
\renewcommand{\[}{\begin{equation}}
\renewcommand{\]}{\end{equation}}

\allowdisplaybreaks
\begin{document}

\pagenumbering{roman}

\begin{titlepage}

\newgeometry{vmargin={15mm}, hmargin={16mm,16mm}}
% \newgeometry{left=1cm, right=1cm, bottom=0.1cm}

\baselineskip=15.5pt \thispagestyle{empty}
\begin{flushright}

\end{flushright}
\vspace{-0.5cm}

\begin{center}
\begin{minipage}{1.0\textwidth}
\centering
{\fontsize{20}{0}\selectfont \bfseries  } ~\\[12pt] 
{\fontsize{15.5}{0}\selectfont \bfseries Fisher Forecast of Finite-Size Effects with Future Gravitational Wave Detectors}
 
\end{minipage}
\end{center}

% \vskip 4pt

\vspace{0.001cm}

\begin{center}
\begin{minipage}{1\textwidth}
\centering
{\fontsize{13}{0} \selectfont Joshua Shterenberg and Zihan Zhou  } 
\end{minipage}
\end{center}

% \vspace{0.01cm}

\begin{center}
% \vskip4pt
\begin{minipage}[c]{1.0\textwidth}
\centering

%\textsl{\fontsize{\affsize}{0}\selectfont $^1$ School of Natural Sciences, Institute for Advanced Study, Princeton, NJ 08540, USA}

\textsl{\fontsize{\affsize}{0}\selectfont  Department of Physics, Princeton University, Princeton, NJ 08540, USA}

\end{minipage}
\end{center}

\vspace{0.3cm}

\begin{center}
\begin{minipage}{0.88\textwidth}
\hrule \vspace{10pt}
\noindent {\bf Abstract}\\[0.1cm]

We use Fisher information theory to forecast the bounds on the finite-size effects of astrophysical compact objects with next-generation gravitational wave detectors, including the ground-based Cosmic Explorer (CE) and Einstein Telescope (ET), as well as the space-based Laser Interferomet Space Antenna (LISA). Exploiting the worldline effective field theory (EFT) formalism, we first characterize three types of quadrupole finite-size effects: the spin-induced quadrupole moments, the conservative tidal deformations, and the tidal heating. We then derive the corresponding contributions to the gravitational waveform phases for binary compact objects in aligned-spin quasi-circular orbits. We separately estimate the constraints on these finite-size effects for black holes using the power spectral densities (PSDs) of the CE+ET detector network and LISA observations. For the CE+ET network, we find that the bounds on the mass-weighted spin-independent dissipation number $\mathcal{H}_0$ are of the order $\mathcal{O}(1)$, while the bounds on the mass-weighted tidal Love number $\tilde{\Lambda}$ are of the order $\mathcal{O}(10)$. For high-spin binary black holes with dimensionless spin $\chi \simeq 0.8$, the bounds on the symmetric spin-induced quadrupole moment $\kappa_s$ are of the order $\mathcal{O}(10^{-1})$. LISA observations of supermassive black hole mergers offer slightly tighter constraints on all three finite-size parameters. Additionally, we perform a Fisher analysis for a binary neutron star merger within the CE+ET network. The bounds on the tidal parameter $\mathcal{H}_0$ and on $\tilde \Lambda$ are around two orders of magnitude better than the current LIGO-Virgo-KAGRA (LVK) bounds. 
 
\vskip15pt
\hrule
\vskip10pt
\end{minipage}
\end{center}

\begin{center}
% \vskip4pt
\begin{minipage}[l]{0.88\textwidth}
%\centering
Email: \href{mailto:}{\texttt{jshteren@princeton.edu}}, \href{mailto:}{\texttt{zihanz@princeton.edu}}
\end{minipage}
\end{center}

\restoregeometry

\end{titlepage}

\thispagestyle{empty}
\setcounter{page}{2}
\tableofcontents

\newpage
\pagenumbering{arabic}
\setcounter{page}{1}

\clearpage
%%%%%%%%%%%%%%%%
\section{Introduction}

The advent of gravitational-wave (GW) astronomy, following the successful detection of GWs by the LIGO-Virgo-KAGRA (LVK) collaboration \cite{LIGOScientific:2018mvr, LIGOScientific:2020ibl, LIGOScientific:2021usb, LIGOScientific:2021djp, Venumadhav:2019tad, Venumadhav:2019lyq, Olsen:2022pin, Nitz:2018imz, Nitz:2019hdf, Nitz:2021uxj, Nitz:2021zwj, Chia:2023tle, Mehta:2023zlk, Wadekar:2023gea}, has significantly heightened global interest in this field over the past decade. Given the ever-increasing sensitivities of GW detectors, precise and accurate waveform modeling is crucial for deepening our understanding of the structure of compact objects \cite{Blanchet:2013haa,Blumlein:2021txe,Blumlein:2021txj,Cho:2022syn,Blanchet:2023soy,Trestini:2023ssa,Blanchet:2024loi,Purrer:2019jcp,Yelikar:2024rmh}. In the early inspiral phase of mergers of compact objects, where the relative velocities of the orbiting bodies remain small, binary systems can be perturbatively described using the methods of the post-Newtonian (PN) expansion (see \cite{poisson2014gravity,Blanchet:2013haa,Blanchet:2024mnz} for comprehensive reviews). In this framework, the binary system is initially modeled as two point particles orbiting around each other. To account for finite-size effects, one goes beyond the point-particle approximation by introducing corrections via the standard multipole expansion. At the quadrupolar level, finite-size effects in GW observables can be broadly categorized into three types: spin-induced multipole moments \cite{Poisson:1997ha,Porto:2005ac,Marsat:2014xea, Levi:2014gsa,Levi:2015msa, Krishnendu:2017shb, Krishnendu:2018nqa,Chia:2020psj, Chia:2022rwc, Lyu:2023zxv}, conservative tidal deformability  \cite{love, Goldberger:2004jt, Flanagan:2007ix, Li:2007qu, Damour:2009vw, Binnington:2009bb, Vines:2011ud, Cardoso:2017cfl}, and tidal heating \cite{hartle_heating, Poisson:1994yf, Tagoshi:1997jy, Alvi:2001mx, Hughes:2001jr, Poisson:2009di, Zahn:2008fk, Ogilvie:2014dwa, poisson_will_2014, murray1999solar}. In this paper, we analyze the capability of future GW detectors, such as Einstein Telescope (ET) \cite{Punturo:2010zz,Hild:2010id}, Cosmic Explorer (CE) \cite{Reitze:2019iox,Evans:2021gyd}, and the Laser Interferometer Space Antenna (LISA) \cite{amaro2017laser} to constrain these quadrupole finite-size effects. The former two will be treated as the CE+ET network. 

In PN theories, the spin-induced quadrupole moment of a self-gravitating body arises from its rotation \cite{Hansen:1974zz,Thorne:1980ru,Poisson:1997ha}. From the standpoint of PN counting, the spin-induced quadrupole moments of the binary system $\{\kappa_1, \kappa_2 \}$ are the dominant finite-size effects. They first appear in the phase of binary waveforms at the 2PN order \cite{Poisson:1997ha}. For Kerr black holes (BHs), the theoretical prediction is $\kappa_1 = \kappa_2 = 1$ \cite{Hansen:1974zz,Thorne:1980ru}. The first sub-leading finite-size effect is the tidal heating \cite{hartle_heating, Poisson:1994yf, Tagoshi:1997jy, Alvi:2001mx, Hughes:2001jr, Poisson:2009di, Zahn:2008fk, Ogilvie:2014dwa, poisson_will_2014, murray1999solar,Saketh:2022xjb,Saketh:2023bul,Chia:2024bwc,Saketh:2024juq} --- also referred to as the tidal dissipation and characterized by the dissipation numbers $\{H_{1\omega}^E, H_{2\omega}^E\}$-- which appears at 2.5PN order for rotating objects and at 4PN order for spherically symmetric objects \cite{Saketh:2022xjb,Saketh:2023bul,Chia:2024bwc,Saketh:2024juq}. Tidal dissipation quantifies the viscous properties of compact objects by describing the irreversible transfer of energy and angular momentum from the surrounding tidal environment into the body itself. A well-known example of this process is observed in the Earth-Moon system \cite{poisson2014gravity, murray1999solar, Endlich:2015mke}. The conservative tidal deformation parameters, which first appear at 5PN order, describe the change in the density distribution and shape of a body under the influence of an external gravitational field. These deformations are characterized by the well-known ``Love numbers" $\{\Lambda_1, \Lambda_2\}$ \cite{Damour:2009vw,Binnington:2009bb,Kol:2011vg,Hui:2020xxx,LeTiec:2020spy,Chia:2020yla,Charalambous:2021mea,Charalambous:2021kcz,Hui:2021vcv,Charalambous:2022rre,Ivanov:2022qqt,Hui:2022vbh,DeLuca:2022tkm,Chia:2023tle,Charalambous:2023jgq,Rodriguez:2023xjd,DeLuca:2023mio,Charalambous:2024tdj,Berti:2024moe}. For Kerr BHs, the Love numbers are identically zero, i.e. $\Lambda_1 = \Lambda_2 = 0$ \cite{Chia:2020yla,Charalambous:2021mea,Charalambous:2021kcz,Hui:2021vcv,Charalambous:2022rre,Ivanov:2022qqt,Saketh:2023bul}. In contrast, for neutron stars (NSs), the Love numbers provide critical information that can be used to distinguish between various degrees of compactness and different equations of state (EoS) \cite{Flanagan:2007ix,Hinderer:2007mb,LIGOScientific:2017vwq,LIGOScientific:2018hze,LIGOScientific:2018cki}, offering insight into the internal structure and composition of these compact objects.

The constraints on the spin-induced quadrupole moments and the tidal Love numbers have been studied extensively in current GW events from LVK's observations \cite{Krishnendu:2017shb,Krishnendu:2018nqa,Kastha:2018bcr,Krishnendu:2019ebd,Krishnendu:2019tjp,Narikawa:2021pak,Saleem:2021vph,LIGOScientific:2021sio}. For BHs, the symmetric spin-induced quadrupole moment parameter $\kappa_s$ is constrained to $|\kappa_s| \lesssim \mathcal{O}(10^2)$ for individual events, with improvements of $|\kappa_s| \lesssim \mathcal{O}(10)$ at the population level \cite{Narikawa:2021pak,Saleem:2021vph}. Similarly, the symmetric mass-weighted tidal Love number $\tilde \Lambda$ is constrained to $|\tilde \Lambda| \lesssim \mathcal{O}(10^4)$, which is consistent with the prediction of a vanishing Love number from general relativity (GR), though far from the precision test \cite{Narikawa:2021pak,Chia:2023tle}.
Tidal heating effects of relativistic compact objects have recently been explored in data analysis \cite{Ripley:2023lsq,Chia:2024bwc,HegadeKR:2024slr}. In Ref.~\cite{Chia:2024bwc}, it is shown that the symmetric mass-weighted dissipation number $\mathcal{H}_0$ for BHs can be constrained to $|\mathcal{H}_0| \lesssim 20$ at the population level. The analysis of finite-size effects for NSs requires more work because of the variety of their compactness and EoS. For spinning neutron stars with $|\chi| \lesssim 0.6$ \cite{LIGOScientific:2018hze}, studies have shown that the spin-induced quadrupole moments $\kappa$ can vary from $2\sim 10$ depending on the EoS \cite{Laarakkers:1997hb,Pappas:2012ns}. The tidal Love number and dissipation also vary widely, from $\mathcal{O}(10^2)$ to $\mathcal{O}(10^4)$, again depending on the compactness and the EoS \cite{HegadeKR:2024agt,Saketh:2024juq}. Parameter estimation for both conservative and dissipative tidal effects has been applied to real data from the binary neutron star (BNS) event GW170817, yielding constraints of $\tilde \Lambda = 300^{+420}_{-230}$ and $\mathcal{H}_0 < 1200$ at the 90\% credible level \cite{LIGOScientific:2018hze,Ripley:2023lsq,HegadeKR:2024slr}. 

Looking ahead, the sensitivities of next-generation gravitational wave detectors such as the ET, CE, and LISA are expected to increase dramatically compared to current detectors. ET and CE are projected to detect compact binaries in the mass range of stellar-mass BHs to roughly one hundred stellar-mass BHs, with sensitivities increased by nearly two orders of magnitude compared with the current LVK observations \cite{Punturo:2010zz,Hild:2010id,Reitze:2019iox,Evans:2021gyd,Srivastava:2022slt,Branchesi:2023mws}. 
This improvement will naturally lead to higher signal-to-noise ratios (SNRs) and tighter constraints on the finite-size effects of compact objects.
LISA, on the other hand, is designed to detect gravitational waves in the millihertz range, which will enable the observation of supermassive ($10^4 \sim 10^7 M_{\odot}$) BBH mergers \cite{amaro2017laser,DuttaRoy:2024qbl}. These systems often involve the merger of a supermassive BH with much smaller compact objects, forming extreme-mass-ratio inspirals (EMRIs). Due to their long inspiral phases, EMRIs provide an exceptional opportunity to test GR and constrain finite-size effects in the strong-field regime.
%with respective SNRs on the order of 100,000 at least . 
For these future detectors, some studies have already assessed the ability of carrying out such tests of GR for EMRIs and other scenarios \cite{amaro2017laser,DuttaRoy:2024qbl}. Together with the future advanced LIGO \cite{Tse:2019wcy,LIGOScientific:2014pky,LIGOScientific:2014pky}, advanced Virgo \cite{VIRGO:2014yos}, LIGO-India \cite{Unnikrishnan:2013qwa,Saleem:2021iwi} and more, the bounds on the finite-size effects of compact objects are going to be rapidly improved both at the individual and population level.

In this paper, we follow the foundations set up by Ref.~\cite{Chia:2024bwc} and extend them to study the signature of finite-size effects of compact objects with future detectors. More specifically, we will adopt the worldline effective field theory (EFT) formalism to model the finite-size effects of compact objects and estimate the bounds on these parameters with future GW detectors. In the EFT framework \cite{Goldberger:2004jt,
Goldberger:2005cd,
Goldberger:2009qd,
Porto:2016pyg,
Levi:2018nxp,
Goldberger:2020fot,Saketh:2022xjb, Saketh:2023bul}, all of the information about the finite-size effects is embedded in the composite operator of the quadrupole moments $Q_{ij}$. In general, $Q_{ij}$ is not known, but for spherically symmetric objects within a slowly varying external tidal environment we can exploit the time derivative expansion and linear response theory to parameterize the quadrupole moment as the following (to first order in $D_{\tau}E_{ij}$):
\begin{equation}
    Q_{ij}^{E{\rm (tidal)}} = - m (G m)^4 \Bigg[\Lambda^E E_{ij} - (G m) H_{\omega}^E \frac{D}{D\tau} E_{ij} \Bigg] ~,
\end{equation}
where $\Lambda^E$ is the Love number and $H_\omega^E$ is the spin-independent dissipation number. The subscript $E$ denotes the parity-even electric-type tidal effects. Going beyond Newtonian gravity, one must also account for parity-odd magnetic-type tidal effects. 
%The dissipation number of Kerr BHs is known from BHPT, as we describe below. Just as with the Love number, the dissipation number for neutron stars provides insight into the internal dynamics and EoSs therein. Previous work (\cite{Ripley:2023lsq}) identifies the dissipation number for NSs to be intimately linked to the effective bulk and shear viscocities of the star, which they bound at $\tilde{\mathcal{H}}_0\le1175^{+136}_{-145}$ at 90\% confidence for GW170817, which follows the real parameter constraint mentioned above. 
If we extend the theory to include intrinsically rotating objects, there are two such additional contributions to the quadrupole moment \cite{Levi:2015msa,Endlich:2015mke,Steinhoff:2021dsn,Cho:2022syn}:
\begin{equation}
    Q_{ij,S}^{E {\rm (tidal)}} = - m (G m)^4 H_{S}^E \chi \hat{S}^{\langle i}{ }_k E^{k|j\rangle} ~,~ Q_{ij}^{E{\rm (spin)}} = - m (G m \chi)^2 \kappa \hat{S}^i{}_k \hat{S}^k{}_j ~,
\end{equation}
where $H_{S}^E$ is the spin-linear dissipation number, $\kappa$ is the spin-induced quadrupole moment parameter and $\hat{S}_{ij}$ is the unit spin tensor. Previous studies have separately examined the constraining power of future detectors on spin-induced quadrupole moments and tidal Love numbers \cite{Krishnendu:2017shb,Lyu:2023zxv,Pacilio:2021jmq,Iacovelli:2023nbv,Huxford:2023qne}. However, tidal dissipation effects have received less attention. More importantly, no analysis has yet simultaneously considered all three types of finite-size effects. In this work, we aim to fill this gap by analyzing the ability of the three aforementioned future detectors to measure the three symmetric mass-weighted finite-size parameters $\kappa_s$, $\mathcal{H}_0$, and $\tilde \Lambda$ for binary compact objects simultaneously.

We separately estimate the projected bounds for these parameters on the CE+ET detector network for stellar-mass BHs, and on LISA for supermassive BHs. Throughout, we adopt the electric-magnetic duality for binary black holes \cite{teukolsky1973perturbations,teukolsky1974perturbations,Chandrasekhar:1985kt,Goldberger:2005cd,Porto:2007qi, Chia:2020yla, Hui:2020xxx}. Furthermore, for the low-spin events, we use the low-spin superradiance condition to achieve better constraints for dissipation numbers (see \cite{Chia:2024bwc} for detailed discussion). Our marginalized constraints on $\mathcal{H}_0$ are of the same magnitude as the theoretical predictions from GR. For high-spin events, where we do not use the superradiance condition, we get slightly less stringent constraints on the dissipation numbers, but significantly better constraints on the spin-induced quadrupole moments --- about an order of magnitude tighter than the values predicted by GR. 
We additionally perform the Fisher analysis for the binary neutron star with the fiducial values chosen from median values of the GW170817 posterior(we set all fiducial dissipation numbers to zero) and find the 90\% credible bounds on the mass-weighted tidal dissipation number to be $\mathcal{H}_0 = 0^{+1.8}_{-1.8}$, and that of the tidal Love number to be $\tilde \Lambda = 456^{+18}_{-18}$.

\noindent \textbf{Outline:} The remaining structure of this paper is as follows. \S\ref{sec:EFT} gives a short review of the worldline EFT formalism and its modeling of the finite-size effects of compact objects.
We focus on three types of finite-size effects: spin-induced quadrupole moments, conservative tidal deformation, and tidal heating. We then derive our corresponding \texttt{IMRPhenomD+FiniteSize} waveform to capture the imprints of these finite-size effects on GW waveforms. \S\ref{sec:Fisher Matrix} presents our Fisher forecasting of the projected bounds on the three finite-size effects mentioned above. 
In \S\ref{sec:conc}, we first summarize our results. Then we identify possible systematic errors in our waveform modeling and give some outlook on future research directions. The Appendix complements \S\ref{sec:EFT} in further detailing the derivations for the waveform observables used.

\noindent \textbf{Notations and Conventions}: We use the natural units $G = c = 1$ unless otherwise specified. We use the $(-+++)$ metric signature, with Greek letters for covariant indices and Latin letters for indices within local tetrads. We use $m_\ell$ to denote the azimuthal angular momentum to avoid confusion with the component mass $m$. We adopt the following conventions for several convenient mass and spin quantities:
\begin{equation}
    \begin{aligned}
    \label{eq:mass_spin param}
M & :=m_1+m_2 \, \qquad 
\eta  :=m_1 m_2 / M^2 \, \qquad
\delta  :=\left(m_1-m_2\right) / M \\
% X_1 & := m_1/M \\
% X_2 & := m_2 / M \\
\chi_i & :=\mathbf{S}_i / m_i^2 \, \qquad
\boldsymbol{\chi}_s  :=\left(\boldsymbol{\chi}_1+\boldsymbol{\chi}_2\right) / 2 \, \qquad
\boldsymbol{\chi}_a  :=\left(\boldsymbol{\chi}_1-\boldsymbol{\chi}_2\right) / 2 
\end{aligned}
\end{equation}
where $\mathbf{S}_i$ is the component spin angular momentum vector and $\chi_i$ is the dimensionless spin. The mass-weighted symmetrized versions of various finite-size parameters are defined in Eqs.~\eqref{eq:sym params}.

\section{Finite-Size Effects on Gravitational Waves}
\label{sec:EFT}
\subsection{Short Review: EFT Formalism with Finite-Size Effects}

The theoretical basis surrounding our work is the worldline effective field theory (EFT) formalism of gravitational compact objects, which has been extensively studied in the literature \cite{Goldberger:2005cd,Porto:2005ac,Porto:2007qi,Goldberger:2020fot,Goldberger:2022ebt,Goldberger:2022rqf}. The construction of the EFT is based on the multipole expansion approach, where the higher order terms are designed to capture more detailed information about the system. 
The leading order term of the EFT describes the compact objects as point particles. More specifically, 
%The existing dynamical degrees of freedom in the PN scale of interest (defined in the Appendix) are the gravity field $g_{\mu\nu}$ itself, 
the point-particle degrees of freedom are captured by the four velocity $u^{\mu}$ of the worldline. When going beyond this point-particle limit, we use the multipole expansion to account for the fine structure within the compact object.  In this paper, we will only focus on the quadrupole terms. Let us denote the co-moving four-tetrads as $e_i^{\mu}; i\in\{0,1,2,3\}$. Within the external gravitational field $g_{\mu\nu}$, one can write down the effective action of the system as \footnote{Note that the convention we use here is different from \cite{Saketh:2023bul,Chia:2024bwc} by a factor of $1/2$.}
\begin{equation}
    S = \int d\tau \Bigg[ - m + \mathcal{L}(Q_{ij}^{E/B}, \dot Q_{ij}^{E/B}) - \frac{1}{2} Q_{ij}^E E^{ij} - \frac{1}{2} Q_{ij}^B B^{ij} \Bigg] ~,
\end{equation}
where $m$ is the mass of the compact object and $Q_{ij}$ is the quadrupole moment. Here, the external electric and magnetic fields $E_{ij}$ and $B_{ij}$ are defined as 
\[
E_{ij}=C_{\mu\rho\nu\sigma}u^{\rho}u^{\sigma} e_{i}^{\mu} e_j^\nu;\qquad B_{ij} = u^\mu e_i^\nu u^\rho e_j^\sigma {}^* C_{\mu\nu\rho\sigma} ~,
\]
where $C_{\mu\nu\rho\sigma}$ is the Weyl tensor of the external gravitational field and ${}^*C_{\mu\nu\rho\sigma}$ stands for its dual. Once we treat $Q_{ij}$ as a dynamical variable, the Lagrangian $\mathcal{L}(Q_{ij}, \dot Q_{ij})$ describes the quadrupole-level internal dynamics of the given particle. Then, to describe the rotating compact objects, we need to recast the tetrads into a co-rotating frame $e_A^\mu$ and identify the angular velocity of the particles:
\begin{equation}
    \Omega^{\mu\nu} \equiv e_A^\mu \frac{D}{D\tau} e^{A\nu} 
\end{equation}
as new dynamical degrees of freedom in the system. The most general action is now extended to be
\begin{equation}
\begin{aligned}
\label{eq:spin EFT action}
    S & = \int d \tau \mathcal{L}\left(u^\mu, \Omega^{\mu\nu},g_{\mu\nu},Q_{ij}^{E/B}, \dot Q_{ij}^{E/B}\right) - \frac{1}{2} \int d\tau \Bigg[ Q_{ij}^E E^{ij} + Q_{ij}^B B^{ij} \Bigg] \\
    & = \int d\tau \Bigg[-m + \frac{I}{2} \Omega_{\mu\nu} \Omega^{\mu\nu} + \mathcal{L}(Q_{ij}^{E/B}, \dot Q_{ij}^{E/B},\Omega^{\mu\nu})\Bigg] - \frac{1}{2} \int d\tau \Bigg[ Q_{ij}^E E^{ij} + Q_{ij}^B B^{ij} \Bigg]
\end{aligned}
\end{equation}
where $I$ is the moment of inertia. As has been demonstrated in \cite{Porto:2005ac,Porto:2008tb,Cho:2022syn}, it is more convenient to adopt the ``Routhian approach" by introducing the conjugate momentum of the angular velocity, i.e. the spin tensors of the particles:
\begin{equation}
    S_{\mu\nu} = -2 \frac{\partial \mathcal{L}}{\partial \Omega^{\mu\nu}} ~.
\end{equation}
We choose the following normalization: $J \equiv \chi G m^2  = \sqrt{1/2 S_{\mu\nu} S^{\mu\nu}}$. For convenience, we further introduce the unit spin tensor $\hat{S}_{\mu\nu} \equiv S_{\mu\nu}/J$ with normalization $\hat{S}_{\mu\nu} \hat S^{\mu\nu} = 2$.
With these definitions and the recasting of the action, we can now clearly identify that all of the finite-size effects in the system are encoded in the composite quadrupole operator $Q_{ij}^{E/B}$.

To further quantify the finite-size effects, we shall use the linear response theory to parameterize the dynamical multipole moments $Q_{ij}$. The contributions can be separated into the tidal part
\begin{equation}
\begin{aligned}
\label{eq:tidal_moment}
    Q^{ij}_{E {\rm (tidal)}} & = -m (G m)^4\left[\Lambda^E E^{ij} -(G m) H_\omega^E \frac{D}{D \tau} E^{i j} + H_S^{E} \chi \hat{S}^{\langle i}{ }_k E^{k|j\rangle}\right] ~, \\
    Q^{ij}_{B {\rm (tidal)}} & = - m (G m)^4 \left[\Lambda^B B^{ij} -(G m) H_\omega^B \frac{D}{D \tau} B^{i j} + H_S^{B} \chi \hat{S}^{\langle i}{ }_k B^{k|j\rangle}\right] ~,
\end{aligned}
\end{equation}
and the spin part
\begin{equation}
\label{eq:spin_momnent}
    Q_{ij{\rm (spin)}}^E = - m (G m \chi)^2 \kappa \hat{S}^i{}_k \hat{S}^k{}_j ~.
\end{equation}
Then, plugging Eq.~\eqref{eq:tidal_moment} and Eq.~\eqref{eq:spin_momnent} into the effective action Eq.~\eqref{eq:spin EFT action}, one can immediately see that the tidal effects are quadratic in curvature and the spin-induced moments are linear in curvature. Furthermore, by analyzing the properties of time-reversal transformations in Eq.~\eqref{eq:tidal_moment}, $\Lambda^{E/B}$ corresponds to the time-reversal even contribution which leads to conservative tidal effects, while $H_\omega^{E/B}$ and $H_S^{E/B}$ are time-reversal odd and correspond to dissipative tidal effects. As a benchmark for our following analysis, we list the fiducial values for the finite-size effects of Kerr BHs extracted from the Kerr metric and linear BH perturbations \cite{Saketh:2022xjb,Saketh:2023bul,Chia:2024bwc}:
\begin{equation}
\label{eq:BH dissipation}
    H_\omega^{E/B} = \frac{16}{45} (1 + \sqrt{1 - \chi^2}) ~,\quad ~ H_S^{E/B} = - \frac{16}{45} (1 + 3 \chi^2) ~,\quad ~ \kappa = 1 ~.
\end{equation}
We also note that, especially when the spins are small, $H_S^{E/B}$ and $H_\omega^{E/B}$ are not independent. These parameters obey the superradiance relation (for more detailed discussion see \cite{Chia:2024bwc})
\begin{equation}
    H_S^{E / B}=-2 \frac{G m \Omega}{\chi} H_\omega^{E/B} ~.
\end{equation}
For small spin Kerr BHs, this simplifies to
\begin{equation}
\label{eq:superradiance condition}
    H_S^{E/B} = - \frac{1}{2} H_\omega^{E/B} ~,
\end{equation}
which can be seen from Eq.~\eqref{eq:BH dissipation}.
For general compact objects, one should consider the electric and magnetic Love/dissipation numbers separately. However, for BHs in four dimensions, these two parameters turn out to have the same values, based on the principle of electric-magnetic duality. As mentioned, we apply this principle for BHs throughout our remaining analysis. We also mention here that the superradiance condition effectively enhance the leading tidal dissipation from 4PN to 2.5PN order and therefore leading to better constraints.

Strictly speaking, there are more spin-dependent finite-size effects
for high-spin systems, such as spin-cubic dissipation numbers, the spin-dependent Love numbers, the spin-induced
octopole moments, and more \cite{Saini:2023gaw,Saketh:2022xjb,Saketh:2023bul}, which we do not consider in this work. These effects may become relevant for the
systems with near extremal BHs that could be detected in the future, for example, from hierarchical BBH mergers.

\subsection{Imprints on Waveforms: \texttt{IMRPhenomD+FiniteSize}}

We now start the discussion of the imprints of these finite-size effects on GW waveforms. As we have mentioned before, we are going to focus on the following finite-size effects for binary systems: spin-induced quadrupole moments $\{\kappa_1 , \kappa_2\}$, static tidal Love numbers $\{\Lambda_1, \Lambda_2\}$ and spin-independent dissipation numbers $\{H_{1\omega}^{E/B}, H_{2\omega}^{E/B}\}$. For small-spin systems, the superradiance condition sets the relationship between the spin-indepdent and spin-linear dissipation numbers $\{H_{1S}^{E/B}, H_{2S}^{E/B}\}$. For binary systems, it is also convenient for us to further define the following mass-weighted symmetric (anti-symmetric) quantities:
\begin{equation}
    \begin{gathered}
    \label{eq:sym params}
\kappa_s \equiv \frac{1}{2} (\kappa_1 + \kappa_2) ~, \quad \kappa_a \equiv \frac{1}{2} (\kappa_1 - \kappa_2)  ~, \\
\mathcal{H}_1^{E / B} \equiv \frac{1}{M^3}\left(m_1^3 H_{1 S}^{E / B}+m_2^3 H_{2 S}^{E / B}\right), \quad \overline{\mathcal{H}}_1^{E / B} \equiv \frac{1}{M^3}\left(m_1^3 H_{1 S}^{E / B}-m_2^3 H_{2 S}^{E / B}\right) \\
\mathcal{H}_0^{E / B} \equiv \frac{1}{M^4}\left(m_1^4 H_{1 \omega}^{E / B}+m_2^4 H_{2 \omega}^{E / B}\right) ~, \quad \tilde{\Lambda} \equiv \frac{16}{13} \frac{\left(m_1+12 m_2\right) m_1^4 \Lambda_1^E}{M^5}+1 \leftrightarrow 2 ~,
\end{gathered}
\end{equation}
where $m_{1},m_2$ are the masses for individual objects and $M = m_1 + m_2$ is the total mass. 

For quasi-circular aligned-spin binary systems, the $(\ell=2,m_\ell = 2)$ gravitational radiation mode takes the following form in the Fourier domain
\begin{equation}
\label{eq:amplitude_phase}
    \tilde{h}(f)=A(f) e^{-i \psi(f)}, \quad \tilde{h}_{+}(f)=\tilde{h}(f) \frac{1+\cos ^2 \iota}{2}, \quad \tilde{h}_{\times}(f)=-i \tilde{h}(f) \cos \iota ~,
\end{equation}
where $A$ is the amplitude, $\psi$ is the phase, $h_{+,\times}$ are the two polarizations of gravitational waves, and $\iota$ is the inclination angle between  the line of sight and the orbital angular momentum. Since we are not considering a specific source in this paper, in \S\ref{sec:Fisher Matrix}, we will marginalize over the inclination angle $\iota$ along with the detector antenna functions when performing the Fisher analysis. 

The evolution of the phase $\psi$ can be derived from the stationary phase approximation \cite{Thorne1980Lectures}. This can be done explicitly by integrating Kepler's third law for the dominant ($
\ell,m_\ell=2,2$) mode of GW emmission:
\[t(v)=t_0+\int dv\frac{1}{\dot v}\qquad\phi(v)=\phi_0+\frac{1}{M}\int dv\frac{v^3}{\dot v}\]
where $\dot v$ can be derived from the energy balance equation given in Eq.~\eqref{eq:ebal}. Iteratively solving Kepler's laws after Taylor expanding about $\dot v$, one can then solve for the phase $\psi(v)=2\pi f t(v)-2\phi(v)$. The contributions involving finite-size effects are then given by the following formula:
\[\psi^{\rm FS}(v)=\frac{3}{128\eta v^5}\left(\psi^{\rm TDN}(v)+\psi^{\rm TLN}(v)+\psi^{\rm SIM}(v)\right)\]
where the tidal dissipation term was first computed in Ref.~\cite{Chia:2024bwc}
\begin{equation}
\begin{aligned}
    \psi^{\rm TDN}&=v^5(1+3\ln v)\left[\frac{25}{8}\mathcal{H}_1^E\chi_s+\frac{25}{8}\bar{\mathcal{H}}_1^E\chi_a\right]\\
    &+v^7\Bigg[\left(\frac{225}{16}\mathcal{H}_1^B+\frac{102975}{896}\mathcal{H}_1^E+\frac{675}{64}\bar{\mathcal{H}}_1^E\delta+\frac{1425}{32}\mathcal{H}_1^E\eta\right)\chi_s\\
    &\qquad+\left(\frac{225}{16}\bar{\mathcal{H}}_1^B+\frac{102975}{896}\bar{\mathcal{H}}_1^E+\frac{675}{64}\mathcal{H}_1^E\delta+\frac{1425}{32}\bar{\mathcal{H}}_1^E\eta\right)\chi_a\Bigg]\\
    &+v^8(1-3\ln v)\left[\frac{25}{4}\mathcal{H}_0^E+\cdots\text{(other spin-dependent terms)}\right] ~.
    \label{eq:TDN phase}
\end{aligned}
\end{equation}
At the $4{\rm PN}$ order, we only focus on the leading symmetric spin-independent tidal dissipation number $\mathcal{H}_0^E$. The other spin-dependent terms at $4{\rm PN}$ are dropped because they are small compared to the spin-dependent ones at $2.5 {\rm PN}$. The contribution from the tidal Love number is given by \cite{Flanagan:2007ix}
\begin{equation}
\label{eq:TLN phase}
        \psi^{\rm TLN} = v^{10}\left[-\frac{39}{2}\tilde\Lambda\right] ~.
\end{equation}
The GW phase from spin-induced moments reads \cite{Krishnendu:2017shb,Cho:2022syn}
\begin{equation}
\label{eq:SIM phase}
    \psi^{\rm SIM} = v^4 \psi^{\rm SIM}_{\rm 2PN} + v^6 \psi^{\rm SIM}_{\rm 3PN} + v^7 \psi^{\rm SIM}_{\rm 3.5PN} ~,
\end{equation}
where the 2PN term is given by
\begin{equation}
    \psi^{\rm SIM}_{\rm 2PN} = -\left(\left(50\delta\kappa_a+50(1-2\eta)\kappa_s\right)\left(\chi_s^2+\chi_a^2\right)+\left(100\delta\kappa_s+100(1-2\eta)\kappa_a\right)\chi_s\chi_a\right) ~,
\end{equation}
the 3PN term is
\begin{equation}
\begin{aligned}
    \psi^{\rm SIM}_{\rm 3PN} & = \Bigg(\left(\frac{26015}{14}-\frac{88510}{21}\eta-480\eta^2\right)\kappa_a+\delta\left(\frac{26015}{14}-\frac{1495}{3}\eta\right)\kappa_s\Bigg)\chi_s\chi_a\\
    &\qquad+\left(\left(\frac{26015}{28}-\frac{44255}{21}\eta-240\eta^2\right)\kappa_s+\delta\left(\frac{26015}{28}-\frac{1495}{6}\eta\right)\kappa_a\right)(\chi_s^2+\chi_a^2) ~,
\end{aligned}
\end{equation}
and the 3.5PN term is
\begin{equation}
\begin{aligned}
    \psi^{\rm SIM}_{\rm 3.5PN} & = (-400\pi\delta\kappa_a-400\pi\kappa_s+\eta(800\pi\kappa_s))(\chi_a^2+\chi_s^2) + (-800\pi\kappa_a+1600\pi\eta\kappa_s-800\pi\delta\kappa_s)\chi_a\chi_s\\
    &\qquad+\Bigg(\left(\frac{3110}{3}-\frac{10250}{3}\eta+40\eta^2\right)\kappa_s+ \left(\frac{3110}{3}-\frac{4030}{3}\eta\right)\delta\kappa_a\Bigg)\chi_s^3\\
    &\qquad+\Bigg(\left(\frac{3110}{3}-\frac{8470}{3}\eta\right)\kappa_a+\left(\frac{3110}{3}-750\eta\right)\delta \kappa_s\Bigg)\chi_a^3\\
    &\qquad+\Bigg(\left(3110-\frac{28970}{3}\eta+80\eta^2\right)\kappa_a+\left(3110-\frac{10310}{3}\eta\right)\delta \kappa_s\Bigg)\chi_s^2\chi_a\\
    &\qquad+\Bigg(\left(3110-\frac{27190}{3}\eta+40\eta^2\right)\kappa_s+\left(3110-\frac{8530}{3}\eta\right)\delta \kappa_s\Bigg)\chi_a^2\chi_s ~.
\end{aligned}
\end{equation}
The mass and spin quantities $\eta, \delta, \chi_s$ and $\chi_a$ are defined in Eq.~\eqref{eq:mass_spin param}. We further incorporate these finite-size effects into the well-known \texttt{IMRPhenomD} waveform for BBH mergers and \texttt{IMRPhenomD}\_\texttt{NRTidalv2} for BNS waveforms. To do this, we introduce our modified waveform: 
\begin{equation}
    \psi(f)=\left\{\begin{array}{l}
\psi^{\mathrm{IMRPhenomD}}(f)+\psi^{\mathrm{FS}}(f)-\psi^{\mathrm{FS}}\left(f_{22}^{\text {ref }}\right), \quad f \leq f_{22}^{\text {tape }} ~, \\
\psi^{\mathrm{IMRPhenomD}}(f)+\psi^{\mathrm{FS}}\left(f_{22}^{\text {tape }}\right)-\psi^{\mathrm{FS}}\left(f_{22}^{\text {ref }}\right), \quad f>f_{22}^{\text {tape }} ~.
\end{array}\right.
\end{equation}
Because the above finite-size GW phase is only valid in the inspiral phase of the binary evolution, it should be terminated when close to merger. To incorporate this, we introduce the so-called tapering frequency $f_{22}^{\rm tape}=\alpha f_{22}^{\rm peak}$, where $f_{22}^{\rm peak}$ is the frequency at the largest amplitude of the waveform of the $(\ell,m_\ell)=(2,2)$ mode. In this paper, we choose $\alpha=0.35$, aligning with the test GR analysis in LVK observations \cite{Mehta:2022pcn}. The reference frequency is the frequency at which the phase of the (2,2) mode of the waveform vanishes, which therefore acts as an overall constant. We denote our new waveform as \texttt{IMRPhenomD+FiniteSize}. For BNS systems, the contribution from the tidal Love numbers has already been incorporated in the known \texttt{IMRPhenomD}\_\texttt{NRTidalv2} waveform, and therefore we only need to add the contributions from tidal dissipation and spin-induced moments to produce a modified version of this waveform. For BBHs, several simplifications can be made using both the electric-magnetic duality (as the electric and the magnetic components are equivalent), and the superradiance condition in Eq.~\eqref{eq:superradiance condition} for low-spin systems.

\section{Fisher Matrix Forecasting}
\label{sec:Fisher Matrix}

In this section, we implement the above \texttt{IMRPhenomD+FiniteSize} GW waveforms and use them to forecast the detection capabilities on finite-size parameters of compact objects using the Fisher information matrix method. %Then we invert the matrix to identify the variance (constraints) on each parameter individually. 
For binary systems, our analysis focuses on the dominant finite-size effects that can be measured accurately: the symmetric spin-induced quadrupole moment $\kappa_s$, the symmetric mass-weighted dissipation number $\mathcal{H}_0$, and the symmetric mass-weighted Love number $\tilde \Lambda$ defined in Eq.~\eqref{eq:sym params}. 
Throughout the analysis, we also make use of the \texttt{GWFast} program \cite{Iacovelli:2022bbs,Iacovelli:2022mbg}.

\subsection{Fisher Information Matrix Basics}

Before we present the concrete results for the bounds on finite-size parameters, we first recap the basics of the Fisher information matrix method \cite{PhysRevD.49.2658,Vallisneri:2007ev,Rodriguez:2013mla,Iacovelli:2022bbs}. In the frequency domain, the observed data $d$ in a detector is a pure waveform $h$ of some parameters $\boldsymbol{\theta}$ overlayed with some known noise function $n$, i.e. $d(f) = \tilde h(f) + n(f)$. The noise function in a single detector is characterized by its one-sided power-spectral density (PSD) $S_n(f)$. Now, instead of calculating the exact likelihood $\mathcal{L}(d|\boldsymbol{\theta})$ of data given some parameters $\boldsymbol{\theta}$, we approximate it by a multivariable Gaussian distribution around certain chosen fiducial values. For that purpose, we Taylor expand the waveform $h(f,\boldsymbol{\theta})$ around this set of fiducial values $\boldsymbol{\theta}_0$ representing the best-fit parameters to linear order
\begin{equation}
    \tilde h(f,\boldsymbol{\theta})=\tilde{h}_0+\tilde{h}_i \delta \theta^i+\ldots ~,
\end{equation}
where $\delta\theta^i \equiv \theta^i - \theta_0^i$ is the first order deviation of the parameters $\theta_i$ from the fiducial choices, and $\tilde{h}_i \equiv \partial_{\theta^i} \tilde h$ is the corresponding deviation of the waveform under parameters' deviation. Within this approximation, the likelihood can be written as
\begin{equation}
\label{eq:likelihood1}
    \mathcal{L}(d \mid \boldsymbol{\theta}) \propto \exp \left[-\frac{1}{2}(n \mid n)+\delta \theta^i\left(n \mid \tilde h_i\right)-\frac{1}{2} \delta \theta^i \delta \theta^j\left(\tilde h_i \mid \tilde h_j\right)\right] ~,
\end{equation}
where the noise-weighted inner product $(\cdot | \cdot) $ is defined as 
\begin{equation}
    (a|b) = 4 {\rm Re} \int_0^\infty df \frac{a^*(f) b(f)}{S_n(f)} ~.
\end{equation}
From a Bayesian point of view, we can treat the likelihood in Eq.~\eqref{eq:likelihood1} as the probability distribution for $\boldsymbol{\delta \theta}$ and we can rewrite the likelihood as
\begin{equation}
\label{eq:likelihood2}
    \mathcal{L}(d|\boldsymbol{\theta}) \propto \exp\Bigg[ - \frac{1}{2} \Gamma_{ij} (\delta\theta^i - \langle \delta \theta^i \rangle)(\delta \theta^j - \langle \delta \theta^j \rangle )\Bigg]
\end{equation}
where the Fisher matrix $\Gamma_{ij}$ is given by
\begin{equation}
\Gamma_{i j}= \Bigg(\frac{\partial \tilde h}{\partial \theta^i} \Bigg| \frac{\partial \tilde h}{\partial \theta^j}\Bigg) = 4 {\rm Re}\int_{f_{\rm low}}^{f_{\rm high}}df\frac{\partial_{\theta^i}\Tilde{h}^*(f,\boldsymbol\theta)\partial_{\theta^j}\Tilde{h}(f,\boldsymbol\theta)}{S_n(f)} ~,
\end{equation}
where the low frequency cutoff is detector dependent. In this paper, we choose the cutoff for CE at 5Hz, ET at 5Hz, and LISA at $10^{-5}$Hz. If one wants to perform the inspiral-only analysis as has been done in \S~4.2 in Ref.~\cite{Chia:2024bwc}, it is sufficient to choose the cutoff at $f_{22}^{\text{tape}}$. From Eq.~\eqref{eq:likelihood2}, we can immediately see that the inverse of the Fisher matrix gives the covariance matrix of the set of parameters $\text{Cov}[\theta_i,\theta_j]$. Therefore the calculation of the Fisher matrix alone is sufficient to determine the variances (and covariances) of the observed parameter values as compared to their fiducial values. 

Given a single detector, the strain of the gravitational wave alone can be written as
\begin{equation}
    \tilde h_{\rm det}(f) = F_{+}(\theta,\phi) \tilde h_+(f) + F_\times (\theta,\phi) \tilde h_\times(f) ~,
\end{equation}
where $\tilde h_+$ and $\tilde h_\times$ are the plus and cross polarization components given in Eq.~\eqref{eq:amplitude_phase}, and $\theta,\phi$ are the sky locations. For this agnostic analysis, we do not fix the sky locations $\theta,\phi$. Instead, we will average over $\theta,\phi$ along with the inclination angle $\iota$, which then leads to \cite{Maggiore:2007ulw}
\begin{equation}
    \tilde h_{\rm det}(f) = \frac{2}{5} \tilde h(f)
\end{equation}
for interferometers with arms perpendicular to each other. For the triangle shape detectors like ${\rm ET}$, this is equivalent to setting \cite{DuttaRoy:2024qbl}
\begin{equation}
    \tilde h_{\rm det}(f) = \frac{\sqrt{3}}{2} \tilde h(f) ~.
\end{equation}
It worth noting that the PSD of LISA in the original review already accounts for the $60^{\circ}$ angle between the detector arms and therefore we only need to add a factor of $\sqrt{4/5}$ in the amplitude of the waveform \cite{DuttaRoy:2024qbl}. This averaging ensures that the strain on the detector is independent of the antenna functions, and will also therefore be independent of all extrinsic parameters about the data we choose, which should be the case for the future detectors.

\subsection{Bounds on Finite-size Parameters}

\subsubsection*{CE+ET Network}

\begin{figure}[h!]
    \centering
    \includegraphics[width=0.8\linewidth]{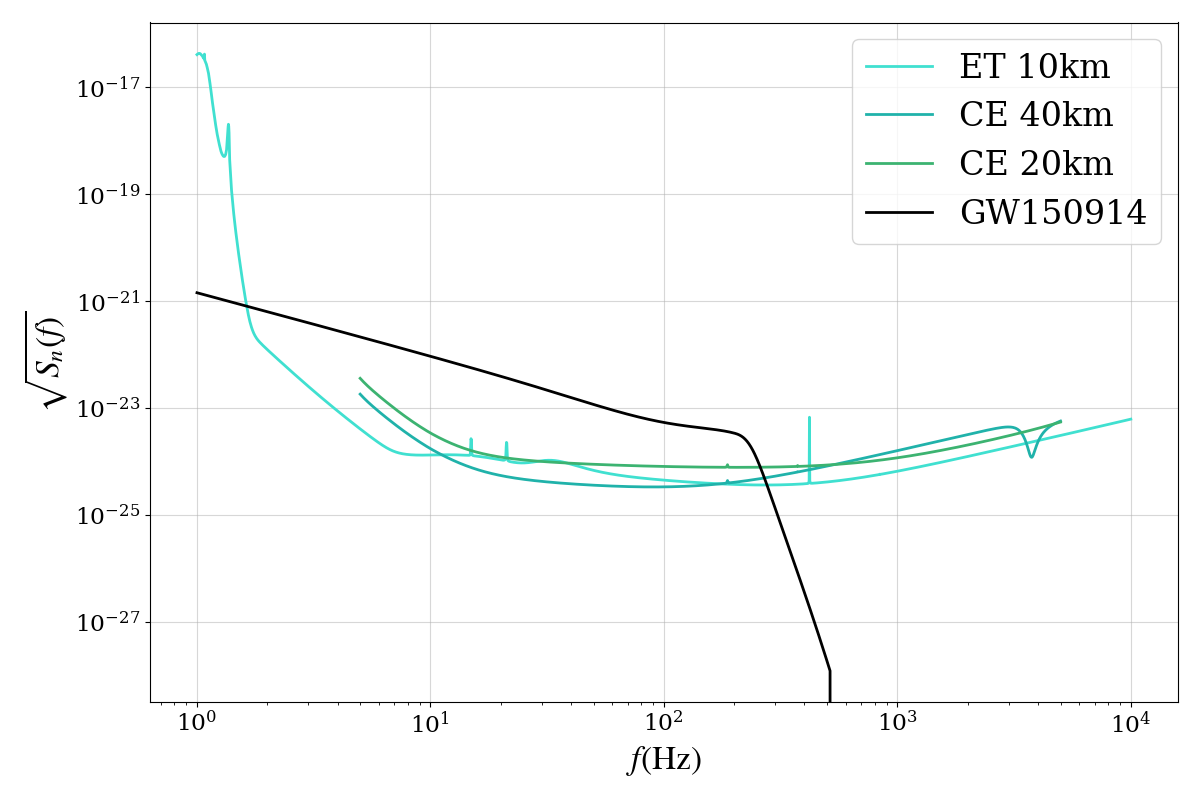}
    \caption{The PSDs of the CE and ET detectors used, plotted with the strain of GW190514-like event with parameters given in the third row (\blue{blue} row) in Table~\ref{table:data}.}
    \label{fig:psds}
\end{figure}

In this section, we show the 90\% credible bounds on the finite-size parameters $\kappa_s, \mathcal{H}_0$ and $\tilde \Lambda$ for BBHs similar to the known events ${\rm GW150914}$ (representative of high mass events) \cite{LIGOScientific:2016aoc}, ${\rm GW151226}$ (representative of low mass events) \cite{LIGOScientific:2016sjg} and BNS similar to ${\rm GW170817}$ \cite{LIGOScientific:2017vwq} using the future detector network ${\rm CE+ET}$: the triangle configuration of ET with $10$ km arms, and two CE detectors with 40km and 20km arms respectively. 
For illustrative purposes, we show the PSDs for the above three detectors in Fig.~\ref{fig:psds} along with the \texttt{IMRPhenomD} waveform of GW150914-like events with parameters given in the second line in Table~\ref{table:data}. The specific PSDs we use are taken from their respective design reviews, for CE \cite{Srivastava:2022slt} and ET \cite{Branchesi:2023mws}. Comparing with the sensitivity curves from the LVK observations, CE and ET improve upon existing detector sensitivities by around two orders of magnitude.  
One common feature of the three chosen events is that they all have relatively small spins, which makes it difficult to constrain the spin-dependent parameters. Therefore, we also consider BBH systems with artificially amplified spins: $\chi_1 = \chi_2 = 0.8$. For these choices of fiducial values, we find much better constraints on the spin-induced quadrupole moment parameter $\kappa_s$. However, the price we pay for this choice is that we lose the constraints from the small-spin superradiance condition in Eq.~\eqref{eq:superradiance condition}, which leads to slightly worse constraints on $\mathcal{H}_0$ for these than for small-spin systems.

\begin{figure}[h!]
    \centering
    \includegraphics[width=\linewidth]{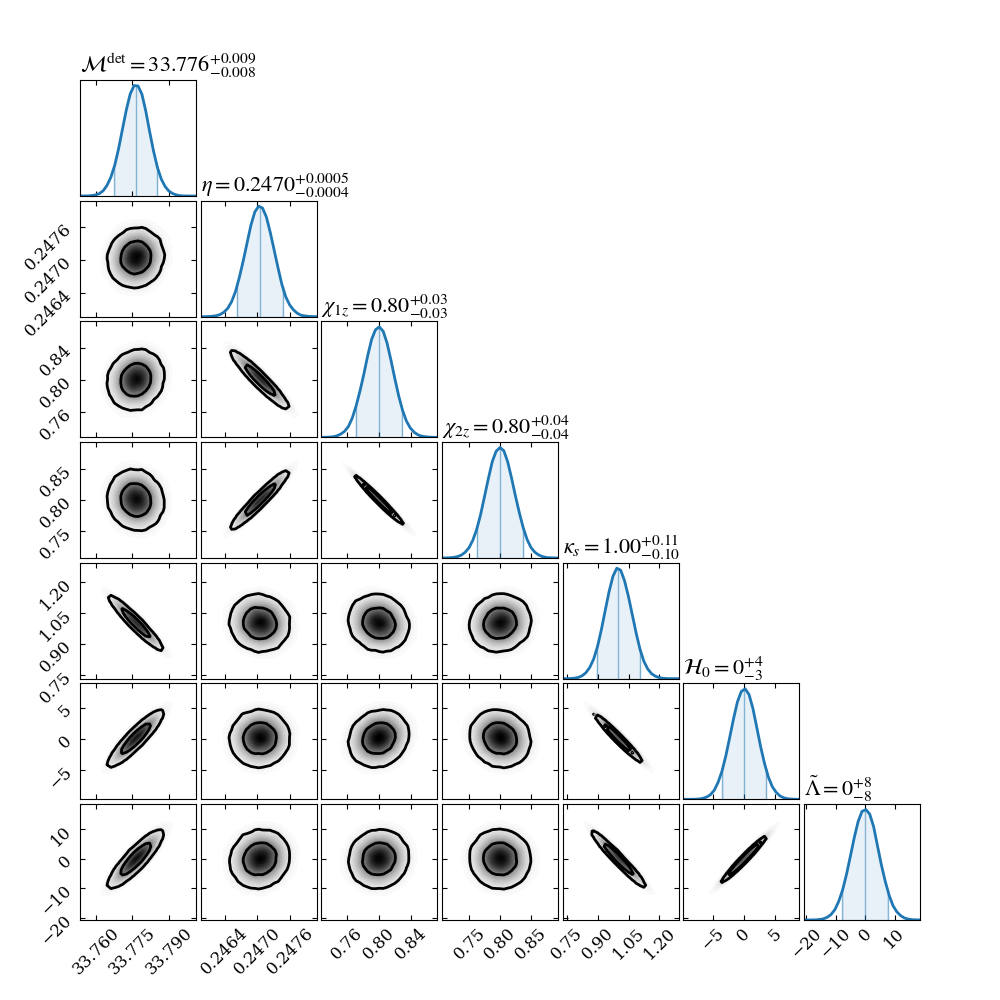}
    \caption{A sample corner plot of the covariance matrix, generated from the Fisher matrix for the GW150914-like event. The center of each graph represents it's mean (fiducial) value, with the left and right shading bounds representing the $1\sigma$ bound. Here, $\mathcal{M}^{\rm det}$ (in unit of $M_{\odot}$) is the chirp mass in the detector frame.}
    \label{fig:cplot}
\end{figure}

In Fig.~\ref{fig:cplot}, we present the Fisher posterior for a GW150914-like event using the PSDs of CE+ET. This cornerplot showcases the relative degeneracies of the parameters that we are trying to bound as well as their individual variances. We find that the constraints on the finite-size parameters are generally strongly correlated with each other. The direction of the degeneracies in the graph can be mostly understood from the waveform phases given in Eqs.~\eqref{eq:TDN phase},~\eqref{eq:TLN phase} and \eqref{eq:SIM phase}. Heuristically, we can collect the first few relevant finite-size effects:
\begin{multline}
    \psi^{\rm FS} \supset \frac{3}{128 \eta v^5} \Bigg[- v^4 \left[50 (1-2\eta) \kappa_s (\chi_s^2 + \chi_a^2) + \cdots\right] + v^8\left[(1-3\log v) \frac{25}{2} \mathcal{H}_0^E + \cdots\right] - v^{10} \left[\frac{39}{2} \tilde \Lambda \right]\Bigg] ~.
\end{multline}
From this expression, we see that the lines of constant phase between 2PN spin-induced moments $\kappa_s$ and 5PN Love number $\tilde \Lambda$ appear to be negatively correlated. Similar arguments also work for the postive correlation between $\mathcal{H}_0^E$ and $\tilde \Lambda$. The opposite correlations between $\kappa_s - \tilde \Lambda$ and $\mathcal{H}_0 - \tilde \Lambda$ ensure the Love numbers to be well-constrained.

Finally, we present the marginalized bounds of finite-size parameters $\kappa_s$, ${\mathcal{H}}_{0}$ and $\tilde \Lambda$ in Table~\ref{table:data}. The first line represents the bounds on finite-size effects of GW170817-like event with the fiducial values for Love numbers $\Lambda_1 = 368.2,\Lambda_2 = 586.5$ chosen from median values of the GW170817 posterior from LVK
observations and dissipation numbers $H_{1\omega}^E =0, H_{2\omega}^E = 0$. Given the relatively small magnitudes of individual spins, we are not able to provide bounds on the spin-induced moments. For tidal dissipations, we choose the fiducial value based on the assumption that neutron stars have almost zero viscosity. In Ref.~\cite{Saketh:2024juq}, the authors have shown that for realistic EoS coming from the relativistic mean-field approximation, the 4PN dissipation number purely comes from the contribution of shear viscosity, which scales as $H_{\omega}^E \propto T^{-2}$, with $T$ being the neutron star core temperature, so long as the inspiral frequency does not hit the NS gravity mode resonance frequency. For relatively low temperature NSs in a binary system (core temperature $T \sim 10^5 K$), the dissipation number ranges from $\mathcal{O}(10^2)$ to $\mathcal{O}(10^4)$, falling sharply with compactness $H_{\omega}^E \propto C^{-6}$, which is defined as $C \equiv G M /R$. Based on the bounds we provide in Table~\ref{table:data}, these low-temperature BNS systems may be visible to these forthcoming detectors in the near future. 

From the second to the last line in Table~\ref{table:data}, we present bounds on finite-size parameters for BBHs. The second and the third lines show the bounds on the GW150914-like event which has relatively high mass. Such an event has a shorter inspiral phase and therefore the constraints are slightly worse than those from the lighter GW151226-like event shown in the fourth and fifth lines. The bounds on dissipation numbers and Love numbers from ET+CE are two order of magnitude better than the bounds we get from the current LVK observations. However, we notice that it is still not possible to rule out the zero dissipaition at the level of individual event which has been claimed for some exotic compact objects \cite{Maselli:2017cmm, Datta:2019euh, Datta:2019epe, Datta:2024vll,Mukherjee:2022wws} . Therefore, to test the nature of BHs, population analysis will be needed. For high spin events, we further put constraints on $\kappa_s$. Since the spin-induced quadrupole moments first appear at 2PN order, the constraints are better than the tidal dissipation and the Love numbers. Finally, the LVK-like events analyzed here are representative high-SNR cases, which may not constitute the majority of detections by CE and ET due to the volume suppression at low redshifts.

\begin{table}
\centering
\begin{tabular}{|p{1cm}|p{1cm}|p{1cm}|p{1.3cm}|p{1.3cm}|p{1.1cm}||p{1cm}|p{1.5cm}|p{1.1cm}||p{1cm}|}
 \hline
 \multirow{3}[0]{*}{} &
 \multicolumn{9}{c|}{90\% bounds for CE+ET Detector Network Results} \\
 \cline{2-10}
 & \multicolumn{5}{c||}{point-particle parameters} & 
 \multicolumn{3}{c||}{finite-size parameters} & \multicolumn{1}{c|}{SNR} \\
 \cline{2-10}
 & $m_1$ & $m_2$ & $\chi_1$ & $\chi_2$ & $z$ & $\kappa_s$ & ${\mathcal{H}}_{0}$&$\tilde \Lambda$ & S/N \\
 \hline
 BNS & 1.496&1.243&0.00513&0.00323&0.0098&$-$&$0^{+1.8}_{-1.8}$&$456^{+18}_{-18}$ & 2479 \\
 \hline
 %1.496&1.243&0.95&0.95&0.0098&$1^{+6}_{-6}$&$0^{+1.1}_{-1.1}$&$0^{+0.08}_{-0.08}$&$10^{-3}$\\
 BBH &36.2&29.1&0.20&0.20&0.094&$-$&$0.1^{+0.7}_{-0.7}$&$0^{+4.9}_{-4.9}$ & 3303 \\
 & \blue{36.2}&\blue{29.1}&\blue{0.80}&\blue{0.80}&\blue{0.094}&\blue{$1^{+0.18}_{-0.16}$}&\blue{$0.0^{+6.6}_{-4.9}$}&\blue{$0^{+13}_{-13}$} & 3689\\
 & 14.2&7.5&0.21&0.21&0.090&$-$&$0.2^{+0.3}_{-0.5}$&$0^{+3}_{-3}$ & 1377\\
 & 14.2&7.5&0.80&0.80&0.090&$1^{+0.06}_{-0.06}$&$0.0^{+3.3}_{-3.3}$&$0^{+8}_{-8}$ & 1443 \\
 \hline
\end{tabular}
\caption{
Data for parameter bounding for CE+ET network. The first row represents a GW170817-like event, the second/third rows represent a low/high spin GW150914-like event, and the fourth/fifth rows represent a low/high spin GW151226-like event. The third row (\blue{blue row}) of this table corresponds to the parameters in Fig \ref{fig:cplot}. Note that this table shows the 90\% bounds, but the corner plot in Fig.~\ref{fig:cplot} shows the $1\sigma$ bounds for display purpose.}
\label{table:data}
\end{table}

\subsubsection*{LISA}

Likewise, we now present similar marginalized bounds for the same three parameters $\kappa_s$, $\mathcal{H}_0$, and $\tilde\Lambda$ for the LISA detector network in Table~\ref{table:datalisa}. The details of calculation remain the same as for the CE+ET network described above. The PSD that we use for LISA analysis is similarly sourced from its respective design review \cite{Robson:2018ifk}. 

The GW signals from stellar-mass events fall outside of the parameter space of what LISA is anticipated to be able to observe. Instead, LISA is targeted to detect the GW signals from supermassive BBHs with mass range from $\mathcal{O}(10^4)$ to $\mathcal{O}(10^7)$ solar masses. We design four such events, with large ($m_1/m_2$=9) and small ($m_1/m_2\approx 1$) mass ratios and large ($\chi_i=0.8$) and small ($\chi_i=0.2$) spins. The specific data we use and the bounds placed on finite-size parameters thereof are recorded in Table~\ref{table:datalisa}. This data shows similar patterns to the CE+ET data, with slightly better bounds overall. The bounds on the spin-induced moments are observed to be comparable with a previous study \cite{Krishnendu:2019ebd}.

\begin{table}
\centering
\begin{tabular}{|p{1.6cm}|p{1.6cm}|p{1.2cm}|p{1.2cm}|p{1.1cm}||p{1.1cm}|p{1.4cm}|p{1.3cm}||p{1.1cm}|}
 \hline
 \multicolumn{9}{|c|}{90\% bounds for LISA Network Results} \\
 \hline
 \multicolumn{5}{|c||}{point particle parameters} & 
 \multicolumn{3}{c||}{finite-size parameters} & \multicolumn{1}{c|}{SNR}\\
 \hline
 $m_1$ & $m_2$ & $\chi_1$ & $\chi_2$ & $z$ & $\kappa_s$ & ${\mathcal{H}}_{0}$&$\tilde\Lambda$ & S/N \\
 \hline
 %$1.496\times10^6$&$1.243\times10^6$&0.00513&0.00323&0.0098&$-$&$0^{+0.05}_{-0.05}$&$0^{+0.003}_{-0.003}$ \\
 %$1.496\times10^6$&$1.243\times10^6$&0.95&0.95&0.0098&$1^{+0.03}_{-0.03}$&$0^{+0.023}_{-0.023}$&$0^{+0.0002}_{-0.0002}$ \\
 $5.5\times10^{5}$&$4.5\times10^{5}$&0.20&0.20&0.512&$-$&$0.11^{+0.13}_{-0.15}$&$0.0^{+1.3}_{-1.3}$ & 24137 \\
 $5.5\times10^{5}$&$4.5\times10^{5}$&0.80&0.80&0.512&$1^{+0.031}_{-0.031}$&$0.1^{+1.3}_{-1.3}$&$0^{+3.3}_{-3.3}$ & 29193 \\
 $9.0\times10^5$&$1.0\times10^5$&0.20&0.20&0.512&$-$&$0.5^{+0.33}_{-0.33}$&$0^{+3}_{-3}$ & 9654 \\
 $9.0\times10^5$&$1.0\times10^5$&0.80&0.80&0.512&$1^{+0.031}_{-0.031}$&$0.4^{+3.1}_{-3.1}$&$0^{+5}_{-5}$ & 14168 \\
 \hline
\end{tabular}
\caption{Data for parameter bounding for LISA. These point particle parameters are not based on any real data but, mirroring the above table, we sample low- and high-spin components and low- and high-mass-ratio events. We believe this data describes a range of characteristic events that LISA will be able to observe.}
\label{table:datalisa}
\end{table}

\section{Conclusions and Outlook}
\label{sec:conc}

In this paper we have utilized our newly constructed \texttt{IMRPhenomD+FiniteSize} waveform to forecast the constraining power of three future detectors on the $\kappa_s$, $\mathcal{H}_0$, and $\tilde\Lambda$ parameters. Making use of the worldline EFT, we have calculated the finite-size modifications to the point-particle PN framework, and have derived the updated waveform. Using the Fisher matrix method on PSDs for CE, ET, and LISA, we were able to indicate various constraining powers on the finite-size effects for BBHs and BNSs. 
For CE+ET, we have found bounds for $\kappa_s$, $\mathcal{H}_0$, and $\tilde\Lambda$ of order $\mathcal{O}(10^{-1})$, $\mathcal{O}(10^0)$, and $\mathcal{O}(10^1)$, respectively, and we identify that LISA better constrains the values of all of these bounds. 

Our work can be extended in various directions that may help gain better understanding of finite-size effects in GW observables. The biggest obstacle this work currently faces is the lack of information about the finite-size contributions to the merger and ringdown phases of the waveform, because these phases exist outside of the scope of the EFT used for our analysis. Additionally, because we use the tapering frequency technique to truncate the GW phase evolution before the merger, we are limiting the SNR of our analysis, and further introducing systematic error in our waveform modeling. For future events with high SNR, we need a much more robust and rigorous treatment to have better control on the finite-size effects on the merger and ringdown phases. Additionally, in the analysis for BNSs, we do not dedicate an analysis to the magnetic-type finite-size effects. Although Ref.~\cite{Binnington:2009bb} has pointed out that the magnetic tidal parameters are much smaller compared with corresponding electric ones, it is still of much interest to quantify the bounds on these parameters. Additionally, Ref.~\cite{Yu:2022fzw} has shown that the non-linear fluid effects can enhance the GW phase by $10\% \sim 20\%$ at GW frequency 1000 Hz even at Newtonian order. Thus, a complete treatment of non-linear tidal effects seems to be necessary for future detectors.

\begin{acknowledgments}

\vskip 8pt

We would like to thank Horng Sheng Chia for suggesting this project. We also thank Muddu Saketh and Matias Zaldarriaga for useful discussions and comments. This work makes use of the \texttt{GWFast} program \cite{Iacovelli:2022bbs,Iacovelli:2022mbg}.

\end{acknowledgments}

%\pagebreak

\appendix

\section{Derivation of \texttt{IMRPhenomD+FiniteSize} Waveform Observables}

In this Appendix, we detail the PN derivation of waveform observables including our modifications. For the purpose of keeping this paper as self-contained as possible, we review existing derivations for the spin contributions, but include the novel components that lead to the phase calculation in section \ref{sec:EFT}.

In the slow inspiral phase (where the PN expansion is still valid), all of the waveform ovservables including the time-evolution of the frequency and phase are directly governed by the energy balance equation:
\[\label{eq:ebal}-\mathcal{F}_{\infty}-\dot M=\dot{\mathcal{E}}\]
Where $\mathcal{F}_{\infty}$ is the energy flux at infinity, $\mathcal{E}$ is the binding energy of the system, $M$ is the total mass, and $\dot X$ denotes time derivative. Here, $\mathcal{E}$ and $\mathcal{F}_{\infty}$ must both be functions of the object masses, spins, finite-size parameters, and the PN expansion parameter $v=(\pi GMf)^{1/3}$. 

The result of the binding energy takes the following form:
\[\mathcal{E}(v)=-\frac{M \eta v^2}{2}\left(\mathcal{E}_{\rm NS}(v)+\mathcal{E}_{\rm SO}(v)v^3+\mathcal{E}_{\rm SS}(v)v^4+\mathcal{E}_{\rm SSS}(v)v^7 + \mathcal{E}_{\rm Love}(v) v^{10}\right)\]
Here $v$ is the expansion velocity. The non-spinning (NS) and spin-orbital (SO) terms are well known and documented in the literature \cite{Blanchet:2024loi, Blanchet:2024mnz,Chia:2024bwc}. The spin quadratic (SS) and cubic (SSS) terms contain contributions both from point-particle terms and spin-induced quadrupole moments. Here, we only list the contribution from spin-induced quadrupole moments as

\begin{align}
    \mathcal{E}_{\rm SS}^{\rm SIM}(v)&=\chi_a\chi_s\left(-\frac{\delta^2\kappa_a}{2}-\delta\kappa_s-\frac{\kappa_a}{2}\right)\nonumber\\
    &+\chi_s^2\left(-\frac{\delta^2\kappa_s}{4}-\frac{\delta\kappa_a}{2}-\frac{\kappa_s}{4}\right)+\chi_a^2\left(-\frac{\delta\kappa_a}{2}+\eta\kappa_s-\frac{\kappa_s}{2}\right)\nonumber\\
    &+v^2\Bigg(\chi_a^2\left(\frac{25\delta\eta\kappa_a}{12}-\frac{35\delta\kappa_a}{12}-\frac{5\eta^2\kappa_s}{6}+\frac{95\eta\kappa_s}{12}-\frac{35\kappa_s}{12}\right)\nonumber\\
    &+\chi_a\chi_s\left(\frac{25\delta\eta\kappa_s}{6}-\frac{35\delta\kappa_s}{6}-\frac{5\eta^2\kappa_a}{3}+\frac{95\eta\kappa_a}{6}-\frac{35\kappa_a}{6}\right)\nonumber\\
    &+\chi_s^2\left(\frac{25\delta\eta\kappa_a}{12}-\frac{35\delta\kappa_a}{12}-\frac{5\eta^2\kappa_s}{6}+\frac{95\eta\kappa_s}{12}-\frac{35\kappa_s}{12}\right)\Bigg)\\
    \mathcal{E}_{\rm SSS}^{\rm SIM}(v)&=\chi_a\chi_s^2\left(-2\delta^2\eta\kappa_a-5\delta^2\kappa_a+6\delta\eta\kappa_s-6\delta\kappa_s-\kappa_a\right)\nonumber\\
    &+\chi_s^3\left(\delta^3(-\kappa_a)-\delta^2\eta\kappa_s-\frac{9\delta^2\kappa_s}{4}-\delta\kappa_a+\frac{\kappa_s}{4}\right)\nonumber\\
    &+\chi_a^2\chi_s\left(-6\delta\kappa_a+4\eta^2\kappa_s+12\eta\kappa_s-6\kappa_s\right)\nonumber\\
    &+\chi_a^3(-2\delta\eta\kappa_s-2\delta\kappa_s+2\eta\kappa_a-2\kappa_a)
\end{align}

%Other terms are calculated in a similar manner, by including other parts of the higher-order effects of the action we described in \S\ref{sec:EFT}. 
Similarly, the energy flux at infinity is known to take a similar form:
\[\mathcal{F}_{\infty}=\frac{32}{5}\eta^2v^{10}\left(\mathcal{F}_{\rm NS}(v)+\mathcal{F}_{\rm SO}(v)v^3+\mathcal{F}_{\rm SS}(v)v^4+\mathcal{F}_{\rm SSS}(v)v^7 + \mathcal{F}_{\rm Love}(v) v^{10} \right)\]
where the NS and SO terms are known in the literature \cite{Blanchet:2024loi, Blanchet:2024mnz,Chia:2024bwc}. The contributions from the spin-induced quadrupole moments to the energy-flux are 

\begin{align}
    \mathcal{F}_{\rm SS}(v)&=\chi_a\chi_s\left(\delta^2\kappa_a+2\delta\kappa_s+\kappa_a\right)+\chi_a^2\left(\delta\kappa_a-2\eta\kappa_s+\kappa_s\right)\nonumber\\
    &+\chi_s^2\left(\delta\kappa_a-2\eta\kappa_s+\kappa_s\right)\nonumber\\
    &+v^2\Bigg(\chi_a^2\left(-\frac{127\delta\eta\kappa_a}{16}+\frac{\delta\kappa_a}{14}+\frac{43\eta^2\kappa_s}{4}-\frac{905\eta\kappa_s}{112}+\frac{\kappa_s}{14}\right)\nonumber\\
    &+\chi_a\chi_s\left(-\frac{127\delta\eta\kappa_s}{8}+\frac{\delta\kappa_s}{7}+\frac{43\eta^2\kappa_a}{2}-\frac{905\eta\kappa_a}{56}+\frac{\kappa_a}{7}\right)\nonumber\\
    &+\chi_s^2\left(-\frac{127\delta\eta\kappa_a}{16}+\frac{\delta\kappa_a}{14}+\frac{43\eta^2\kappa_s}{4}-\frac{905\eta\kappa_s}{112}+\frac{\kappa_s}{14}\right)\Bigg)\nonumber\\
    &+v^3\Bigg(\chi_a^2\left(4\pi\delta\kappa_a-8\pi\eta\kappa_s+4\pi\kappa_s\right)+\chi_a\chi_s\left(8\pi\delta\kappa_s-16\pi\eta\kappa_a+8\pi\kappa_a\right)\nonumber\\
    &+\chi_s^2\left(4\pi\delta\kappa_a-8\pi\eta\kappa_s+4\pi\kappa_s\right)\Bigg)\\
    \mathcal{F}_{\rm SSS}(v)&=\chi_a\chi_s^2\left(\frac{13}{3}\delta^2\eta\kappa_a+\frac{27\delta^2\kappa_a}{16}+\frac{4\delta\eta\kappa_s}{3}+\frac{15\delta\kappa_s}{8}+\frac{3\kappa_a}{16}\right)\nonumber\\
    &+\chi_a^2\chi_s\left(\frac{95\delta\eta\kappa_a}{12}+\frac{15\delta\kappa_a}{8}-\frac{26\eta^2\kappa_s}{3}+\frac{25\eta\kappa_s}{6}+\frac{15\kappa_s}{8}\right)\nonumber\\
    &+\chi_s^3\left(-\frac{7\delta\eta\kappa_a}{4}+\frac{5\delta\kappa_a}{8}-\frac{26\eta^2\kappa_s}{3}-3\eta\kappa_s+\frac{5\kappa_s}{8}\right)\nonumber\\
    &+\chi_a^3\left(\frac{29\delta\eta\kappa_s}{6}+\frac{5\delta\kappa_s}{8}+\frac{43\eta\kappa_a}{12}+\frac{5\kappa_a}{8}\right)
\end{align}

For general tidal dissipation numbers, the horizon flux is given by 
\begin{equation}
    \begin{aligned}
\dot{M}(v)= & \frac{1}{2} \left(9 \overline{\mathcal{H}}_1^E \eta^2 \chi_a+9 \mathcal{H}_1^E \eta^2 \chi_s\right) v^{15}+ \frac{1}{2} \left[\left(9 \mathcal{H}_1^B \eta^2+\frac{45 \mathcal{H}_1^E \eta^2}{2}+\frac{9}{2} \overline{\mathcal{H}}_1^E \delta \eta^2-27 \mathcal{H}_1^E \eta^3\right) \chi_s\right. \\
& \left.+\left(9 \overline{\mathcal{H}}_1^B \eta^2+\frac{45 \overline{\mathcal{H}}_1^E \eta^2}{2}+\frac{9}{2} \mathcal{H}_1^E \delta \eta^2-27 \overline{\mathcal{H}}_1^E \eta^3\right) \chi_a\right] v^{17}+ 9 \mathcal{H}_0 \eta^2 v^{18}
\end{aligned}
\end{equation}
where the dissipation numbers $\mathcal{H}_1^{E/B}, \overline{H}_1^{E/B}$ and $\mathcal{H}_0$ are defined in Eq.~\eqref{eq:sym params}.  The binding energy $\mathcal{E}_{\rm Love}$ and the energy flux $\mathcal{F}_{\rm Love}$ involving Love numbers are given in Ref.~\cite{Flanagan:2007ix}.

%\newpage
\phantomsection
\addcontentsline{toc}{section}{References}
\bibliographystyle{utphys}
\bibliography{references.bib}

\end{document}